\documentclass[pra]{revtex4}
\usepackage{amsmath}
\usepackage{graphicx}
\begin{document}
\title{Domain Wall Dynamics in Ginzburg-Landau-Type Equations with Conservative Quantities}
\author{Hidetsugu Sakaguchi and Hiroshi Akamine}
\affiliation{Department of Applied Science for Electronics and Materials,
Interdisciplinary Graduate School of Engineering Sciences, Kyushu
University, Kasuga, Fukuoka 816-8580, Japan}
\begin{abstract}
In the Ginzburg-Landau equation, there are domain walls connecting two metastable states. The dynamics of domain walls has been intensively studied, but there remain still unsolved but crucial problems even for a single domain. We study the domain wall dynamics in three different Ginzburg-Landau-type equations satisfying conservation laws. In a modified $\phi^4$ model satisfying the law of energy conservation and the Lorentz invariance, the motion of a domain wall is accelerated and the velocity approaches its maximum. In a one-dimensional model of eutectic growth, the order parameter is conserved and a domain wall connecting a metastable uniform state and a spatially periodic pattern appears.  We try to find a selection rule for the wavelength  of a spatially periodic pattern. In a model equation for martensitic transformation, a domain wall connecting a uniform metastable state and a zigzag structure appears which propagates at a high velocity.\end{abstract}
\maketitle
\section{Introduction}
 The time-dependent Ginzburg-Landau equation has been intensively studied to elucidate the dynamics of phase transitions~\cite{rf:1}. In the first-order phase transition, there is a parameter range in which metastable and stable phases coexist. A domain wall that connects bistable states plays an important role in phase transition dynamics~\cite{rf:2}. 
A standard one-dimensional time-dependent Ginzburg-Landau equation is written as 
\begin{equation}
\frac{\partial \phi}{\partial t}=-\frac{\delta U({\phi})}{\delta \phi},
\end{equation}
where $\phi$ denotes an order parameter for the phase transition and $U$ is a functional corresponding to the free energy. If $U$ is assumed to be 
\begin{equation}
U=\int\left \{-\frac{1}{2}\phi^2+\frac{1}{4}\phi^4-\frac{\alpha}{3}\phi^3+\frac{1}{2}\left (\frac{\partial \phi}{\partial x}\right )^2\right \}dx.
\end{equation}
Eq.~(1) is written as 
\begin{equation}
\frac{\partial \phi}{\partial t}=\phi-\phi^3+\alpha \phi^2+\frac{\partial^2 \phi}{\partial x^2},
\end{equation}
where $\alpha$ is a control parameter. Here, the diffusion constant is assumed to be 1 for simplicity. 
This is a time-dependent Ginzburg-Landau equation. 
When $\alpha=0$, $\phi=\pm 1$ are bistable states, and the domain that connects these two states is stationary. If $\alpha>0$, the bistable uniform states are expressed as 
$\phi=\alpha/2\pm \sqrt{1+\alpha^2/4}$. The state $\phi=\alpha/2+\sqrt{1+\alpha^2/4}$ has a lower free energy. The domain wall that connects the two states moves at a constant velocity in the direction as the state of the lower energy becomes dominant. The domain wall solution is approximated at $\phi=(\alpha/2)+A{\rm tanh}\{k(x-ct)\}$ with $A=\sqrt{1+\alpha^2/4}$. The integration between $-\infty$ and $\infty$ after multiplying Eq.~(3) by $\partial \phi/\partial x$ yields
\begin{equation}
-c\int_{-\infty}^{\infty}\left(\frac{\partial \phi}{\partial x}\right )^2dx=\int_{-\infty}^{\infty}\left (\phi-\phi^3+\alpha \phi^2+\frac{\partial^2 \phi}{\partial x^2}\right )\frac{\partial \phi}{\partial x}.
\end{equation}
Owing to the substitution of the ansatz of $\phi$ into Eq.~(4), the velocity $c$ is evaluated as $c=-(2\alpha/3)/(4kA^2/3)=-\alpha/\sqrt{2}$ for a small $\alpha$. This is a known result but we show a numerical simulation of this equation as an introduction to more complicated systems.  
Figure 1(a) shows the time evolution of the domain wall at $\alpha=0.02$. The initial condition is $\phi=\alpha/2+A{\rm tanh}\{k(x-90)\}$ with $k=A/\sqrt{2}$. The no-flux boundary conditions $\partial \phi/\partial x=0$ are imposed at $x=0$ and $l=100$. 
 Figure 1(b) shows the time evolution of the position $x_0$ (solid line) of the domain wall and the theoretical line by $(-0.02/\sqrt{2})t+90$ (dashed line). Here, the position $x_0$ is evaluated as a point satisfying $\phi(x)=0$. 
\begin{figure}[tbp]
\begin{center}
\includegraphics[height=4.cm]{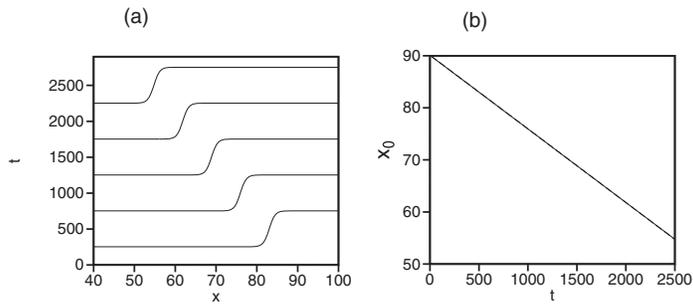}
\end{center}
\caption{(a) Time evolution by Eq.~(3) of a domain wall at $\alpha=0.02$. (b) Time evolution of the position $x_0$ (solid line) of the domain wall and the theoretical line by $(-0.02/\sqrt{2})t+90$ (dashed line). }
\label{fig1}
\end{figure}

Although the dynamics of a single domain wall in a standard Ginzburg-Landau equation is already well understood, there are unsolved problems of the domain wall dynamics in some more complicated Ginzburg-Landau-type equations, which are important in material sciences.  
In this paper, we study the dynamics of a single domain wall in three Ginzburg-Landau-type equations satisfying different conservative quantities. Some of them play an important role in determining the metallographic structure. They have similar forms at first glance, but the dynamics is rather different owing to different conservation laws. The domain wall dynamics is an interesting topic from the viewpoint of nonlinear dynamics.  
\begin{figure}[tbp]
\begin{center}
\includegraphics[height=4.cm]{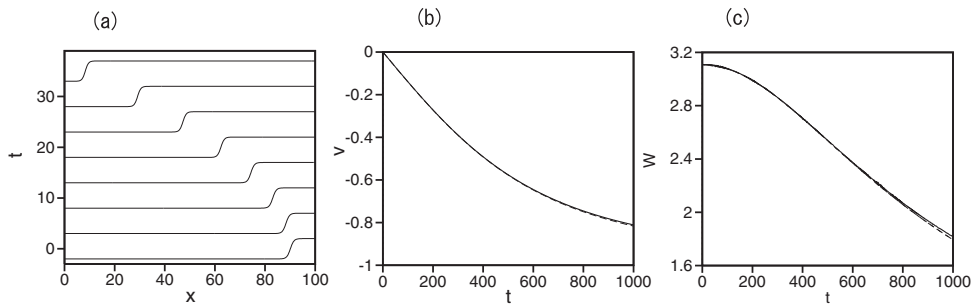}
\end{center}
\caption{(a) Time evolution of a domain wall by Eq.~(5) at $\alpha=0.002$. (b) Time evolutions of the kink velocity $v=d\xi/dt$ by the direct numerical simulation of Eq.~(5) at $\alpha=0.002$ (solid curve) and $v=(-\alpha t)/\sqrt{2+\alpha^2 t^2}$ (dashed curve). (c) Time evolutions of the kink width $W$ by the direct numerical simulation at $\alpha=0.002$ (solid line) and $W$ calculated using Eq.~(12) (dashed line) and Eqs.~(9) and (10) (dotted line). The difference between  dashed and dotted lines is hardly visible. }
\label{fig2}
\end{figure}
\section{Domain Dynamics in a Modified $\phi^4$ Model} 
 In this section, we study the domain wall dynamics in a model system satisfying the energy conservation law and the invariance for the Lorentz transformation.  We consider a system where the potential energy $U$ is expressed as  Eq.~(2) and the total energy is conserved. The variable $\phi$ can be interpreted as a variable such as the lattice deformation.   The model equation is written as  
\begin{equation}
\frac{\partial^2 \phi}{\partial t^2}=\phi-\phi^3+\alpha \phi^2+\frac{\partial^2 \phi}{\partial x^2}.
\end{equation}
This equation is called the $\phi^4$ equation in the case of $\alpha=0$. 
We call Eq.~(5) with a nonzero $\alpha$ a modified $\phi^4$ model in this paper. The total energy $E=\int(1/2)(\partial \phi/\partial t)^2dx+U$ is conserved. This equation can be derived from the Lagrangian
\begin{equation}
L=\int_0^l \left \{\frac{1}{2}\left (\frac{\partial \phi}{\partial t}\right )^2-\frac{1}{2}\left (\frac{\partial \phi}{\partial x}\right )^2+\frac{\phi^2}{2}-\frac{\phi^4}{4}+\frac{\alpha \phi^3}{3}\right\}dx,
\end{equation}
from the Euler-Lagrange equation $\partial/\partial t(\delta L/\delta \phi_t)=\delta L/\delta \phi$. Here, the system size is denoted as $l$. 
An approximate solution for the domain wall or the kink solution is written as 
\begin{equation}
\phi=\frac{\alpha}{2}+A{\rm tanh}\{k(t)(x-\xi(t))\}, 
\end{equation}
where $A=\sqrt{1+\alpha^2/4}$.  
The substitution of Eq.~(7) into Eq.~(6) yields\cite{rf:3}
\begin{equation}
L\sim \frac{2}{3}A^2k\dot{\xi}^2+\frac{1}{2}B\frac{A^2\dot{k}^2}{k^3}-\frac{2}{3}A^2k
+\left\{\frac{1}{4}+\frac{\alpha^2}{4}+\frac{\alpha^4}{24}\right\}l-\frac{A^4}{3k}+\frac{\alpha A^{3}(l-2\xi)}{3},
\end{equation}
where $B=(\pi^2-6)/9$, $\dot{\xi}=\partial\xi/\partial t$ and $\dot{k}=\partial k/\partial t$.  
The Euler-Lagrange equations $\partial/\partial t(\partial L/\partial \dot{\xi})=\partial L/\partial \xi$ and $\partial/\partial t(\partial L/\partial \dot{k})=\partial L/\partial k$ yield
\begin{eqnarray}
\frac{d}{dt}(k\dot{\xi})&=&-\frac{\alpha A}{2},\\
B\left \{k\frac{d^2k}{dt^2}-\frac{3}{2}\left(\frac{dk}{dt}\right )^2\right \}&=&\frac{2}{3}(\dot{\xi}^2-1)k^3+\frac{A^2}{3}k.
\end{eqnarray}
Equations (9) and (10) have the symmetry with respect to the time reversal $t\rightarrow -t$. 
If the approximation $d^2k/dt^2=dk/dt=0$ in Eq.~(10) is assumed, $k=A/\sqrt{2(1-\dot{\xi}^2)}$. From Eq.~(9), $k\dot{\xi}=(-\alpha A/2)t$; therefore, $\dot{\xi}^2/\{2(1-\dot{\xi}^2)\}=\alpha^2t^2/4$. The kink velocity $v$ and $k$ therefore satisfy
\begin{eqnarray}
v&=&\frac{d\xi}{dt}=\frac{-\alpha t}{\sqrt{2+\alpha^2 t^2}},\\
k&=&\frac{\sqrt{1+\alpha^2/4}\sqrt{2+\alpha^2 t^2}}{2}.
\end{eqnarray}
Owing to the symmetry of the model equation, there is a right-propagating kink solution  with a velocity $v=\alpha t/\sqrt{2+\alpha^2 t^2}$. In this case, $A=-\sqrt{1+\alpha^2/4}$, which corresponds to the anti-kink solution.
The domain wall is accelerated when $\alpha\ne 0$ and $|v|$ approaches the maximum velocity of 1. The width of the domain wall is contracted as $1/k$, which is analogous to the Lorentz contraction, because the model equation is invariant for the Lorentz transformation.    
 
We have performed direct numerical simulation using the Runge-Kutta method by discretizing the space with $\Delta x=0.1$ and the time with $\Delta t=0.0005$. 
Figure 2(a) shows the time evolution of the domain structure for $\alpha=0.002$. The initial condition is $\phi=\alpha/2+A{\rm tanh}\{k(x-90)\}$ with $k=A/\sqrt{2}$, and the no-flux boundary conditions $\partial\phi/\partial x=0$ are imposed at $x=0$ and $x=l=100$. It is observed that the structure of the domain wall is rather stable and it is accelerated. A similar type of acceleration was observed in the numerical simulation of the sine-Gordon equation~\cite{rf:4}; however, there were controversies whether or not the domain wall obeys the Newtonian dynamics~\cite{rf:5}.   
We have performed a numerical simulation at the same parameter $\alpha=0.002$ in a larger system of $l=1200$. Figure 2(b) shows a comparison of the time evolution of $v(t)=d\xi(t)/dt$ by a direct numerical simulation with $v=(-\alpha t)/\sqrt{2+\alpha^2 t^2}$ by Eq.~(11). Good agreement is observed, which suggests that the kink dynamics in the modified $\phi^4$ model does not obey the Newtonian dynamics. 
We have performed numerical simulation of Eqs.~(9) and (10) at $\alpha=0.002$. 
The plot of the velocity $\dot{\xi}$ almost completely overlaps with the dashed line $v=(-\alpha t)/\sqrt{2+\alpha^2 t^2}$. This implies that the approximation of $d^2k/dt^2=dk/dt=0$ in Eq.~(10) is rather good. 
Figure 2(c) shows the width $W=x_2-x_2$ (solid line) of the kink solution by the direct numerical simulation, where $x_2$ and $x_1$ are defined as positions satisfying $\phi(x_2)=0.8$ and $\phi(x_1)=-0.8$. The width of the kink solution decreases as a result of the acceleration.  
The dashed line shows the width $W=x_2-x_1$, where $\phi(x_2)=\alpha/2+A{\rm tanh}\{k(t)(x_2-\xi(t))\}=0.8$, $\phi(x_1)=\alpha/2+A{\rm tanh}\{k(t)(x_1-\xi(t))\}=-0.8$, and $k=\sqrt{1+\alpha^2/4}\sqrt{2+\alpha^2 t^2}/2$. The agreement of the solid and dashed curves implies the Lorentz contraction of the kink solution by the acceleration. The dotted line shows the width $W=x_2-x_1$ calculated using $k(t)$ obtained numerically by  Eqs.~(9) and (10). The difference between the dashed line and the dotted line is hardly visible in this plot.   
\begin{figure}[tbp]
\begin{center}
\includegraphics[height=4.cm]{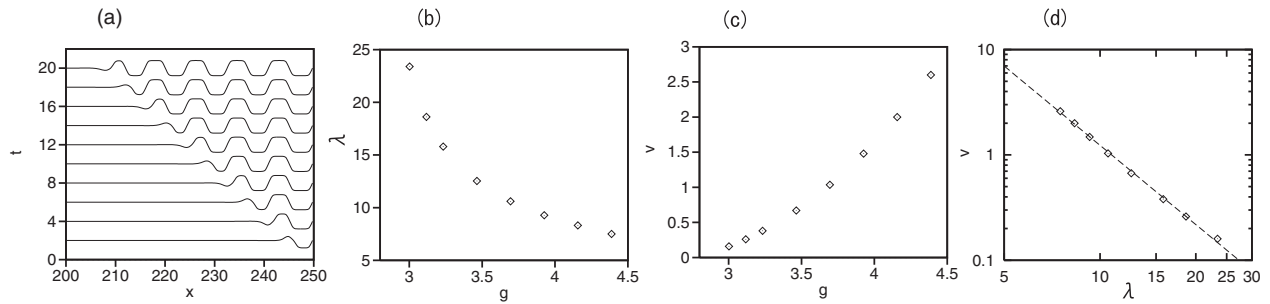}
\end{center}
\caption{(a) Time evolution of a domain wall by Eqs.~(13) and (14) at $g=4.16$. (b) Wavelength $\lambda$ vs $g$. (c) Velocity $v$ vs $g$. (d) Double-logarithmic plot of $\lambda$ vs $v$. The dashed line is the line of $v\propto 1/\lambda^{2.5}$. }
\label{fig3}
\end{figure}
\section{One-Dimensional Model for Eutectic Growth}
In this section, we consider a one-dimensional model equation for eutectic growth where the order parameter is conserved. 
The model equation is written as 
\begin{equation}
\frac{\partial\phi}{\partial t}=-\frac{\partial^2 }{\partial x^2}\left (-\frac{\delta U({\phi})}{\delta \phi}\right ),
\end{equation}
where  
\begin{equation}
U=\int\left \{\frac{1}{2}\phi^2-\frac{g}{4}\phi^4+\frac{1}{6}\phi^6+\frac{1}{2}\left (\frac{\partial \phi}{\partial x}\right )^2\right \}dx.
\end{equation}
The three uniform states of $\phi=0$ and $\phi=\pm\sqrt{(g+\sqrt{g^2-4})/2}$ are all locally stable. When $g>g_c=\sqrt{16/3}$, the uniform state of $\phi=0$ becomes metastable, and the uniform states of $\phi\ne 0$ have a lower free energy. Although the free energy is the lowest for the uniform 
state  $\phi(x)=\pm\sqrt{(g+\sqrt{g^2-4})/2}$ for $g>g_c$,  the total sum of the order parameter $S=\int\phi dx$ is conserved in the time evolution of Eq.~(13).  Then, the state of the lowest free energy is a phase-separated state where $\phi(x)=\sqrt{(g+\sqrt{g^2-4})/2}$ for a sufficiently large $x$ satisfying $x>0$, $\phi(x)=-\sqrt{(g+\sqrt{g^2-4})/2}$ for a sufficiently small $x$ satisfying $x<0$, and there is a domain wall near $x=0$.   
Since the uniform state of $\phi=0$ is locally stable, nucleation is necessary to obtain a state of the lower free energy. The completely phase-separated state does not appear from a natural time evolution after nucleation.  After nucleation, the region of the lower-energy states grows in time, and a spatially periodic structure appears owing to the conservation law of the order parameter. The spatially periodic structure is numerically stable. In principle, it is possible that the wavelength of the spatially periodic structure becomes gradually larger and larger by coarsening toward a completely phase-separated state, because the free energy decreases with coarsening; however, the time evolution is too slow and cannot be observed in the numerical simulation.  This type of crystal growth occurs in the eutectic or eutectoid growth of alloys~\cite{rf:6}. A typical example is pearlite in Fe and C alloy or carbon steel~\cite{rf:7}. A lamellar structure of ferrite and cementite appears as a result of the eutectoid growth.  

We have performed numerical simulation of Eq.~(13) by the pseudospectral method, because the fourth-order spatial derivative is included. 
The boundary conditions are $\partial\phi/\partial x=\partial^3\phi/\partial x^3=0$ at $x=0$ and $l$. 
Figure 3(a) shows that a spatially periodic state appears at $g=4.16$. The initial condition is $\phi=5(x-l)e^{-(l-x)^2}$ with $l=250$. 
A domain wall between the uniform state of $\phi=0$ and a spatially periodic state propagates in the left direction. 
The wavelength of the spatially periodic pattern and the average velocity of the domain wall as a function of $g$ are shown by rhombi in Figs.~3(b) and 3(c) under the initial condition. We have checked that the wavelength and velocity hardly depend on the initial conditions. 
Here, the average velocity is evaluated from the average slope of the plot of the domain wall position and time.   

When $g$ approaches the critical value $g_c$, the wavelength increases and the velocity of the domain wall decreases. The selection rule for the wavelength and  average velocity is not a trivial problem. It is not determined by the principle of the lowest free energy or a principle related to the linear stability.  There is a theory of eutectic growth proposed by Jackson and Hunt~\cite{rf:8}; that is, a theory of the growth in a direction perpendicular to the stripe pattern. In this theory, the velocity is given as a pulling velocity in directional solidification, and the wavelength is determined by the minimum supercooling hypothesis. On the other hand, there is no clear theory regarding the wavelength of the periodic pattern or the average velocity of the domain wall in the one-dimensional model equation Eq.~(13). 
Figure 3(d) shows a logarithmic plot of the relation of $\lambda$ and $v$. The dashed line is the line of $v\propto 1/\lambda^{2.5}$. In the theory of Jackson and Hunt, $v\propto 1/\lambda^2$ is predicted. A similar type of power law is observed in our numerical simulation, but the exponent is slightly different. 

There is no theory regarding the one-dimensional eutectic growth model. We try to approximately evaluate the velocity and wavelength. 
 Equation (13) can be rewritten as
\begin{eqnarray}
\frac{\partial \phi}{\partial t}&=&-\frac{\partial^2 h}{\partial x^2},\\
\frac{\partial^2\phi}{\partial x^2}&=&\phi-g\phi^3+\phi^5+h.
\end{eqnarray}
The uniform state of $\phi=0$ and the spatially periodic state with a wavelength $\lambda$: $\phi_{\lambda}(x)$ are solutions of Eq.~(16) for $h$=0. 
The spatially periodic solution $\phi_{\lambda}(x)$ of Eq.~(16) at $h=0$ satisfies
\begin{equation}
\frac{1}{2}\left (\frac{\partial\phi_{\lambda}}{\partial x}\right )^2=\frac{1}{2}\phi_{\lambda}^2-\frac{g}{4}\phi_{\lambda}^4+\frac{1}{6}\phi_{\lambda}^6+C,
\end{equation}
where $C>0$ is a constant of the integral. Because $u=\phi_{\lambda}^2$ satisfies
\[\frac{du}{\sqrt{2Cu+u^2-gu^3/2+u^4/3}}=2dx,\]
$u$ can be formally expressed with an elliptic function. 
The peak positions of the periodic pattern seem to be almost stationary. We assume that only an envelope function propagates with an average velocity on a stationary periodic pattern; thus the domain wall solution can be expressed as 
\begin{equation}
\phi=f(x+vt-x_0)\phi_{\lambda}(x),
\end{equation}
where $f(x)$ is an envelope function satisfying $f(x)\rightarrow 1$ for $x\rightarrow \infty$ and $f(x)\rightarrow 0$ for $x\rightarrow -\infty$, and $x_0$ is the initial position of the domain wall. Here, we assume $f(x)=\exp(kx)$ for $x<0$ and $f(x)=1$ for $x>0$ as a simple form of the envelope function. 
The time derivative $\partial \phi/\partial t$ can be approximated as $\partial \phi/\partial t=0$ for $x>x_0-vt$ and $\partial \phi/\partial t=kv\phi_{\lambda 0}e^{k(x+vt-x_0)}$ for $x<x_0-vt$, 
where $\phi_{\lambda 0}$ is $\phi_{\lambda}(x)$ at $x=x_0-vt$, if $k$ is sufficiently larger than $1/\lambda$. From Eq.~(15), 
$\partial h/\partial x$ can be approximated as
\begin{eqnarray}
\frac{\partial h}{\partial x}&=&0,\;\;\; {\rm for}\;\; x>x_0-vt,\nonumber\\
\frac{\partial h}{\partial x}&=&-v\phi_{\lambda 0}e^{k(x+vt-x_0)},\;\;\; {\rm for}\;\; x<x_0-vt
\end{eqnarray}
because $\partial h/\partial x=0$ for $x\rightarrow \pm\infty$.  
Then, $h(x)$ can be evaluated as 
\begin{eqnarray}
h(x)&=&0,\;\;\; {\rm for}\;\; x>x_0-vt,\nonumber\\
h(x)&=&-\frac{v\phi_{\lambda 0}}{k}e^{k(x+vt-x_0)},\;\;\; {\rm for}\;\; x<x_0-vt.
\end{eqnarray}
\begin{figure}[tbp]
\begin{center}
\includegraphics[height=4.cm]{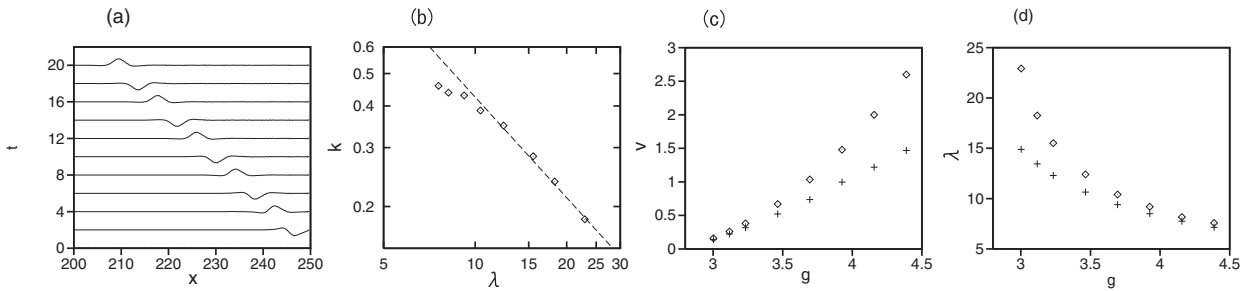}
\end{center}
\caption{(a) Time evolution of $h(x,t)$ obtained by numerical simulation of Eqs.~(15) and (16) at $g=4.16$. (b) Approximate $k$ as a function of $\lambda$ in a double-logarithmic plot. The dashed line is the line of $k=4.25/\lambda$. 
(c) Numerically obtained values (rhombi) of $v$ and approximate values ($+$): $v=2kC/\langle\phi_{\lambda 0}^2\rangle$. (d) Wavelength $\lambda$ (rhombi) obtained by numerical simulation and wavelength ($+$) determined by the maximization of the free-energy decay rate. }
\label{fig4}
\end{figure}
Figure 4(a) shows the time evolution of $h(x,t)$ obtained by numerical simulation at $g=4.16$. A localized pattern propagates in the $-x$ direction, and the sign of the peak value changes periodically in time, which corresponds to the temporal change in $\phi_{\lambda 0}$ in Eq.~(20), although the wave form is different from Eq.~(20).  
The width $\Delta x$ of the localized pattern can be numerically obtained from the relation  
\[(\Delta x)^2=\frac{\int_{-\infty}^{\infty}(x-x_m)h^2dx}{\int_{-\infty}^{\infty}h^2dx},\]
where $x_m=\int_{-\infty}^{\infty}xh^2dx/\int_{-\infty}^{\infty}h^2dx$. The above approximation $f(x)=\exp(kx)$ for $x<0$ yields the width $\Delta x=1/(2k)$. 
Figure 4(b) shows the approximate values of $k$ as a function of $\lambda$ in the double-logarithmic plot. The dashed line is the line of $k=4.25/\lambda$. The width of the domain wall $1/k$ is roughly proportional to $\lambda$ in a parameter range of large $\lambda$ values. 

The integration after multiplying Eq.~(16) by $\partial \phi/\partial x$  yields
\begin{eqnarray}
& &\int_{-\infty}^{\infty}\left [\frac{1}{2}\left (\frac{\partial\phi_{\lambda}}{\partial x}\right )^2-\frac{1}{2}\phi_{\lambda}^2+\frac{g}{4}\phi_{\lambda}^4-\frac{1}{6}\phi_{\lambda}^6\right ]dx=C=\int_{-\infty}^{\infty}h\frac{\partial\phi}{\partial x}dx=-\int_{-\infty}^{\infty}\frac{\partial h}{\partial x}\phi dx\nonumber\\
&\sim &\int_{-\infty}^{x_0-vt}v\phi_{\lambda 0}e^{k(x+vt-x_0)}\phi_{\lambda 0}e^{k(x+vt-x_0)}dx=\frac{v\phi_{\lambda 0}^2}{2k}.
\end{eqnarray}
This is a relation at a certain time, and $\phi_{\lambda 0}$ changes periodically owing to the propagation of the domain wall. We take the average of Eq.~(21) with respect to one wavelength of the spatially periodic function.  Then, we obtain the relation
\[
C=\frac{v\langle\phi_{\lambda 0}^2\rangle}{2k},
\]     
where $\langle\phi_{\lambda 0}^2\rangle$ is the average over the wavelength $\lambda$: $\langle \phi_{\lambda 0}^2\rangle=(1/\lambda)\int_0^{\lambda}\phi_{\lambda 0}^2dx$. The time-averaged velocity $v$ is therefore evaluated approximately at $v=2kC/\langle\phi_{\lambda 0}^2\rangle$. 
Figure 4(c) shows a comparison of the numerically obtained time-averaged velocities (rhombi) with the approximate values $v=2kC/\langle\phi_{\lambda 0}^2\rangle$ ($+$). The theoretical values are smaller than the numerical ones; however, the agreement is acceptable.  

The selection rule of the wavelength of the spatially periodic state is not understood yet. 
We try to evaluate the wavelength using a variant of the maximum entropy production rate principle~\cite{rf:9}.  
 The free energy density of the spatially periodic state of wavelength $\lambda$ is 
\[\Delta U=\frac{1}{\lambda}\int_0^{\lambda}\left \{\frac{1}{2}\phi_{\lambda}^2-\frac{g}{4}\phi_{\lambda}^4+\frac{1}{6}\phi_{\lambda}^6+\frac{1}{2}\left (\frac{\partial\phi_{\lambda}}{\partial x}\right )^2\right \}dx.\]
The free-energy decay rate owing to the domain wall motion can be evaluated as $-v\Delta U$, because the region of the spatially periodic state increases as $vt$.  The free-energy decay rate is a quantity analogous to the entropy production rate. To evaluate $-v\Delta U$, we use rough approximation with $v=2kC/\langle\phi_{\lambda 0}^2\rangle$ and $k=4.25/\lambda$. 
Figure 4(d) shows a comparison of the numerically obtained wavelength $\lambda$ (rhombi) with the one ($+$) determined by the maximization of the free-energy decay rate. The theoretical values are smaller than the numerical values, partly because many approximations are used for the evaluation. However, the maximization principle of the free-energy decay rate is not an established principle, and more investigations are necessary. 

\begin{figure}[tbp]
\begin{center}
\includegraphics[height=4.cm]{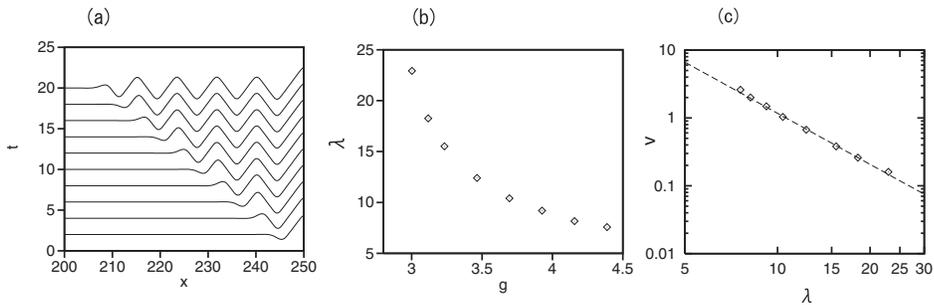}
\end{center}
\caption{(a) Time evolution of a domain wall at $g=4.16$ for Eq.~(23). (b) Wavelength $\lambda$ vs $g$. (c) Double-logarithmic plot of $\lambda$ and $v$. The dashed line denotes the line of $v\propto 1/\lambda^{2.5}$. }
\label{fig5}
\end{figure}
\section{Domain Wall Dynamics in a Model Equation for Martensitic Transformation}
In the shape-memory alloys, the lattice deformation is an order parameter. 
The Ginzburg-Landau theory was proposed for martensitic transformation in shape-memory alloys~\cite{rf:10}.  
In the Ginzburg-Landau theory, the elastic energy is assumed to be
\begin{equation}
U=\int\left \{\frac{1}{2}z_x^2-\frac{g}{4}z_x^4+\frac{1}{6}z_x^6+\frac{1}{2}\left (\frac{\partial^2 z}{\partial x^2}\right )^2\right \}dx,
\end{equation}
where $z$ is the displacement and $z_x=\partial z/\partial x$ denotes the strain.  If a relaxation dynamics is assumed, the time evolution of $z$ is expressed as 
\begin{equation}
\frac{\partial z}{\partial t}=-\frac{\delta U}{\delta z}=\frac{\partial^2z}{\partial x^2}-\frac{\partial^4 z}{\partial x^4}+\frac{\partial}{\partial x}\left \{-g\left (\frac{\partial z}{\partial x}\right )^3+\left (\frac{\partial z}{\partial x}\right )^5\right \}.
\end{equation} 
We have performed direct numerical simulation of Eq.~(23) by the pseudospectral method. 
Figure 5(a) shows the time evolution of $z(x,t)$ obtained by the numerical simulation of Eq.~(23) at $g=4.16$. The initial condition is $z(x)=4e^{-0.2(l-x)^2}$. 
The boundary conditions are $\partial z/\partial z=\partial^3z/\partial x^3=0$ at $x=0$ and $l=250$.
A zigzag structure propagates in the left. Figure 5(b) shows the relation of the wavelength $\lambda$ and $g$. Figure 5(b) is almost the same as Fig.~3(b). 
This is because $z_x$ obeys Eqs.~(13) and (14), which can be shown if Eq.~(23) is differentiated with respect to $x$.  Figure 5(c) shows a double-logarithmic plot of $\lambda$ and $v$. The dashed line denotes the line of $v\propto \lambda^{-2.5}$. The behavior is almost the same as that shown in Fig.~3(d). 

However, martensitic transformation is rather fast and propagates at a velocity on the order of the sound velocity in some materials.  The collective and mechanical motions of lattice deformation occur, and the diffusion of atoms is negligible during martensitic transformation. 
In the fast martensitic transformation, the lattice deformation is expected to obey the energy conservation law. A model equation satisfying the energy conservation law is expressed as 
\begin{equation}
\frac{\partial^2 z}{\partial t^2}=-\frac{\delta U}{\delta z}=\frac{\partial^2z}{\partial x^2}-\frac{\partial^4 z}{\partial x^4}+\frac{\partial}{\partial x}\left\{-g\left (\frac{\partial z}{\partial x}\right )^3+\left (\frac{\partial z}{\partial x}\right )^5\right \}.
\end{equation} 
This equation can be derived from the Euler-Lagrange equation for the Lagrangian $L=\int_0^l \{(1/2)(\partial z/\partial t)^2dx-U$.  
Both the total energy and the total deformation $\int zdx$ are conserved in the time evolution of Eq.~(24). 
As numerical methods of martensitic transformation, the molecular dynamics simulation and a phase field model are used~\cite{rf:11,rf:12}. 
In the phase field model, relaxation dynamics is assumed. 
Molecular dynamics is a model of a large number of discrete atoms or molecules. The nonlinear partial differential equation Eq.~(24) has not yet been studied. 

We have performed numerical simulation of Eq.~(24) by the pseudospectral method. 
Figure 6(a) shows the time evolution of $z(x,t)$ obtained by the numerical simulation of Eq.~(24) at $g=4.16$. The boundary conditions are $\partial z/\partial x=\partial^3z/\partial x^3=0$ at $x=0$ and $l=250$.  The initial condition is $6e^{-0.1(l-x)^2}$. 
A zigzag structure appears. A domain wall structure is rather clear even in this conservative system. The time evolution of the zigzag structure corresponds to the subsequent twinning in martensitic transformation, i.e., a phenomenon that twin crystals are sequentially created. The zigzag structures are interpreted to be twin crystals. In the numerical simulation shown in Fig.~6(a), the zigzag structure is not stationary, but its peak positions move slowly to the left. Furthermore, coarsening to a wider zigzag structure is observed in a region far from the domain wall.  The amplitude of the zigzag structure is rather large as a result of the coarsening near the right end $x\sim l$. The wavelength near the tip position is plotted in Fig.~6(b). We have confirmed that the wavelength near the tip position hardly depends on initial conditions. 
The wavelength is smaller than that in the relaxation dynamics shown in Fig.~5(b), which is close to Fig.~3(b). The time-averaged velocity of the domain wall is plotted in Fig.~6(c). The velocity is rather 
higher than that for the relaxation dynamics, which is close to that in Fig.~3(c). Figure 6(d) shows a double-logarithmic plot of $\lambda$ and $v$. 
It is approximated at $v\propto 1/\lambda^{0.92}$. A power law is observed, but its exponent is rather smaller than the exponent 2.5 for the relaxation dynamics.  The selection rule of the wavelength and velocity for Eq.~(24) remains to be elucidated.  
\begin{figure}[tbp]
\begin{center}
\includegraphics[height=4.cm]{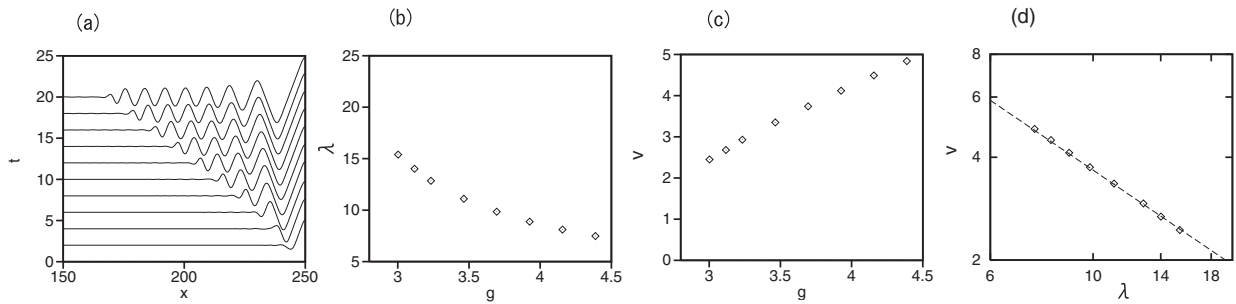}
\end{center}
\caption{(a) Time evolution of a domain wall by Eq.~(24) at $g=4.16$. (b) Wavelength $\lambda$ vs $g$. (c) Velocity $v$ vs $g$. (d) Double-logarithmic plot of $\lambda$ and $v$. The dashed line is the line of $v\propto 1/\lambda^{0.92}$.}
\label{fig6}
\end{figure}
\begin{figure}[tbp]
\begin{center}
\includegraphics[height=4.cm]{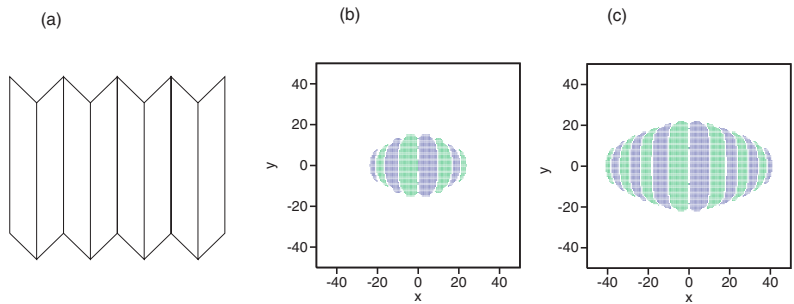}
\end{center}
\caption{(a) Schematic figure of martensitic transformation. (b) Snapshot pattern of a two-dimensional zigzag structure in the time evolution by Eq.~(25) at $t=10$ for $g=4.16, D=2$, and $E=4$. In the blue region, $z_x<-0.5$; in the green region, $z_x>0.5$. (c)  Snapshot pattern of a two-dimensional zigzag structure at $t=15$.  }
\label{fig7}
\end{figure}

Martensitic transformation occurs in three-dimensional crystals. 
The schematic transformation is shown in Fig.~7(a) in two dimensions. 
A zigzag structure in the periphery is induced by martensitic transformation. 
If we consider the formation of the simplest zigzag structure such as that in  Fig.~7(a), instability occurs only in the $x$-direction and the deformation in the $y$-direction does not grow, although a more complex deformation needs to be taken into consideration in general three-dimensional martensite deformation.  
  Then, we can generalize Eq.~(24) to a two-dimensional model as 
\begin{equation}
\frac{\partial^2 z}{\partial t^2}=\frac{\partial^2z}{\partial x^2}+D\frac{\partial^2z}{\partial y^2}-E\frac{\partial^4 z}{\partial x^4}+\frac{\partial}{\partial x}\left \{-g\left (\frac{\partial z}{\partial x}\right )^3+\left (\frac{\partial z}{\partial x}\right )^5\right \}. 
\end{equation} 
Here, the linear elastic force is assumed in the $y$-direction. 
We have performed two-dimensional numerical simulation by the pseudospectral method. 
Figures 7(b) and 7(c) show two snapshot patterns of $z_x$ at $t=10$ and 15 for $g=4.16, D=2$, and $E=4$. The system size is $l\times l=75\times 75$, and the boundary conditions are $\partial z/\partial x=\partial^3z/\partial x^3=0$ at $x=-l/2$ and $x=l/2$ and $\partial z/\partial y=0$ at $y=-l/2$ and $y=l/2$. The initial condition is $z(x,y)=10e^{-0.1(x^2+y^2)}$. 
The zigzag structure evolves in an elliptic manner.     
A similar subsequent twinning was experimentally observed in shape memory alloys~\cite{rf:13}. 
\section{Conclusion}
We have studied the dynamics of domain walls in Ginzburg-Landau-type equations with some conservative quantities. We have observed the acceleration of the domain wall in a modified $\phi^4$ model. In a one-dimensional model of eutectic growth, a domain wall between the uniform state of $\phi=0$ and a spatially periodic pattern appears owing to the conservative law of order parameter. We have tried to approximately evaluate the wavelength of the spatially periodic pattern and the velocity of the domain wall; however, the quantitative argument is left for future study. Finally, we have numerically studied the one- and two-dimensional model equations for martensitic transformation and observed a rather regular subsequent twinning. The elucidation of the selection rule for the wavelength of the stripe patterns is left for future study. 

\end{document}